\documentclass[pdflatex,sn-mathphys-num, twoside]{sn-jnl}% 

\usepackage{graphicx}%
\usepackage{multirow}%
\usepackage{amsmath,amssymb,amsfonts}%
\usepackage{amsthm}%
\usepackage{mathrsfs}%
\usepackage[title]{appendix}%
\usepackage{xcolor}%
\usepackage{textcomp}%
\usepackage{manyfoot}%
\usepackage{booktabs}%
\usepackage{algorithm}%
\usepackage{algorithmicx}%
\usepackage{algpseudocode}%
\usepackage{listings}%
\hypersetup{
  hypertexnames=false,
  pdftitle={Quantum-enabled framework for the Advanced Encryption Standard in the post-quantum era},
  pdfauthor={Albert Nieto Morales, Arit Kumar Bishwas, Joel Jacob Varghese},
  pdfkeywords={quantum random number generator, advanced encryption standard, post-quantum cryptography, cryptographic key management, quantum-enabled AES}
}

\theoremstyle{plain}%
\newtheorem{theorem}{Theorem}%
\newtheorem{proposition}[theorem]{Proposition}% 
\theoremstyle{definition}%
\begin{document}

\title[QRNG-assisted AES-256 key management]{Quantum-enabled framework for the Advanced Encryption Standard in the post-quantum era}

\author[1]{\fnm{Albert} \sur{Nieto Morales}}
\author*[1]{\fnm{Arit Kumar} \sur{Bishwas}}\email{arit.kumar.bishwas@pwc.com}
\author[1]{\fnm{Joel Jacob} \sur{Varghese}}

\affil*[1]{\orgdiv{Innovation Hub}, \orgname{PricewaterhouseCoopers}, \orgaddress{\country{USA}}}

\abstract{The Advanced Encryption Standard (AES)-256 remains resistant to known quantum key-search attacks when keys are uniformly generated. We present quantum-enabled AES (QE-AES), a key-management wrapper that provisions unmodified AES-256 with output delivered by a commercial quantum random-number generator and periodically replaces epoch keys. Evaluation covers Special Publication 800-90B diagnostics on vendor-processed output, capture-backed performance, a live delivery probe, and a simulated correlation-power-analysis model. QE-AES does not enlarge the AES key space or provide masking; it supplies a separate source path and bounds key reuse under an independent-epoch assumption. Entropy delivery, rather than AES, is the principal performance constraint.}

\keywords{quantum random number generator, advanced encryption standard, post-quantum cryptography, cryptographic key management, quantum-enabled AES}

\maketitle

\section{Introduction}\label{sec1}

The Advanced Encryption Standard (AES) is the standardized foundation of modern symmetric encryption \cite{nist2001fips197,daemen2002design}. Grover's algorithm gives a generic quadratic speedup for exhaustive key search \cite{grover1996fast,bonnetain2019quantum}, while compromised key-generation state remains relevant independently of that speedup \cite{nist2020sp800133}. AES-256 retains an estimated \(2^{128}\) generic quantum key-search cost when its keys are uniformly distributed. This work therefore studies how AES-256 is \emph{keyed and rekeyed} in practice rather than modifying the cipher.

Three concerns motivate the design. First, Grover's algorithm reduces generic exhaustive key search for AES-$n$ from \(O(2^n)\) to \(O(2^{n/2})\), motivating the 256-bit variant for a \(2^{128}\)-order generic quantum-search margin \cite{grover1996fast,bonnetain2019quantum}. Second, a cryptographically secure pseudorandom number generator (CSPRNG) remains computationally secure under its assumptions, but compromise, poor seeding, or state recovery can weaken generated keys \cite{vaudenay2005classical}. Third, physical leakage can reveal partial key information and reduce the remaining search space independently of whether the residual search is classical or quantum accelerated \cite{yan2019quantum}.

Our threat model therefore includes a large-scale quantum adversary capable of Grover search, a ``harvest-now-decrypt-later'' (HNDL) strategy \cite{mosca2017cybersecurity}, compromise of the host randomness path, and classical physical leakage such as timing or electromagnetic emissions. It also permits combinations of leakage and quantum-accelerated residual search \cite{krachenfels2021automated}. Rekeying limits the data and traces associated with one epoch key, but it does not prevent leakage within an epoch or replace implementation countermeasures.

While post-quantum cryptography standardization focuses principally on asymmetric algorithms \cite{stebila2016post,nist2020ir8309}, symmetric deployments must also preserve sound key generation and lifecycle controls. This paper studies a quantum-random-number-generator (QRNG) input to the AES-256 key lifecycle. Its contributions are:
\begin{itemize}
    \item an explicit, decryptable key-management specification that leaves the AES-256 cipher core unchanged;
    \item diagnostics of the output delivered by a commercial QRNG, with its vendor post-processing boundary and experimental limitations stated explicitly;
    \item performance measurements that distinguish capture-backed conditioning cost from live-device delivery limits;
    \item a scoped simulation illustrating how a per-key trace budget interacts with rekeying, rather than evaluating masking or a concrete operational rekey interval; and
    \item a conditional confidentiality statement that isolates the effect of key-distribution quality.
\end{itemize}
We also identify integration points for existing cryptographic infrastructure while treating key-generation conformance and deployment-specific validation as requirements rather than assumed properties \cite{nist2020sp800133}.

\textbf{Scope and terminology.}
We use \emph{quantum-enabled AES} (QE-AES) to mean AES-256 whose master and epoch keys are provisioned through a QRNG-based path while its round function, mode of operation, and ciphertext format remain standard. QE-AES is a key-management wrapper, not a new block cipher. We use \emph{quantum-safe} only in the computational sense of resistance under stated assumptions to a quantum-capable adversary making classical oracle queries; it does not imply information-theoretic security. A \emph{quantum side-channel} denotes classical physical leakage combined with quantum-accelerated search over the remaining key space. Accordingly, this study claims neither a new hardness assumption, certification of the inaccessible detector-level source, nor an increase in AES-256's asymptotic Grover resistance.

\section{Background and related work}\label{sec2}

AES is a standardized 128-bit block cipher supporting 128-, 192-, and 256-bit keys \cite{nist2001fips197,daemen2002design}. AES-256 is the reference block cipher for NIST post-quantum security category 5 \cite{nist2020ir8309}. That classification assumes a key drawn from the intended distribution: weak seeding or exposed generator state can undermine a deployment without changing AES itself \cite{nist2020sp800133}. PQC key encapsulation addresses establishment of shared secrets \cite{nist2022pqc,braithwaite2016hybrid}; the present work instead concerns endpoint key provisioning and rotation.

Cryptographic pseudorandom generators expand a secret seed deterministically and are secure under their construction and state-protection assumptions. Physical random-number generators sample nondeterministic processes, while QRNGs use quantum measurements such as photon paths or optical-field quadratures \cite{rarity1994quantum,gabriel2016continuous,herrero2017quantum}. Device-independent designs can provide stronger certification under Bell-test assumptions \cite{pironio2010random,bierhorst2018experiment}, whereas deployable QRNG designs trade assurance scope, rate, and implementation complexity \cite{herrero2017quantum,IDQuantique2025}. The Quantis USB used here exposes only vendor-post-processed bytes rather than detector-level samples; consequently, our experiments characterize that delivered interface and not the underlying physical source.

Prior quantum-AES research has primarily studied circuit implementations or attack cost. Almazrooie et al.\ constructed a 928-qubit AES-128 circuit \cite{almazrooie2018quantum}; Wang et al.\ optimized subkey integration \cite{wang2022quantumcircuit}; Huang et al.\ simplified quantum S-box circuits \cite{huang2022synthesizing}; and Zou et al.\ reduced resources for AES-256 \cite{zou2020quantum}. Sharma et al.\ combined quantum randomness for key generation with classical bulk encryption \cite{sharma2023post}. QE-AES differs by retaining standard AES-256 while specifying the complete key-provisioning and epoch-rekeying path, measuring delivered-stream and integration costs, and expressing confidentiality conditionally on the resulting key distribution. QRNG-based key generation is not itself new and remains subject to NIST entropy-source and key-generation guidance \cite{nist2018sp80090b,nist2020sp800133}.

\section{Methodology}\label{sec3}
\begin{figure}[!ht]
    \centering
    \includegraphics[width=\textwidth]{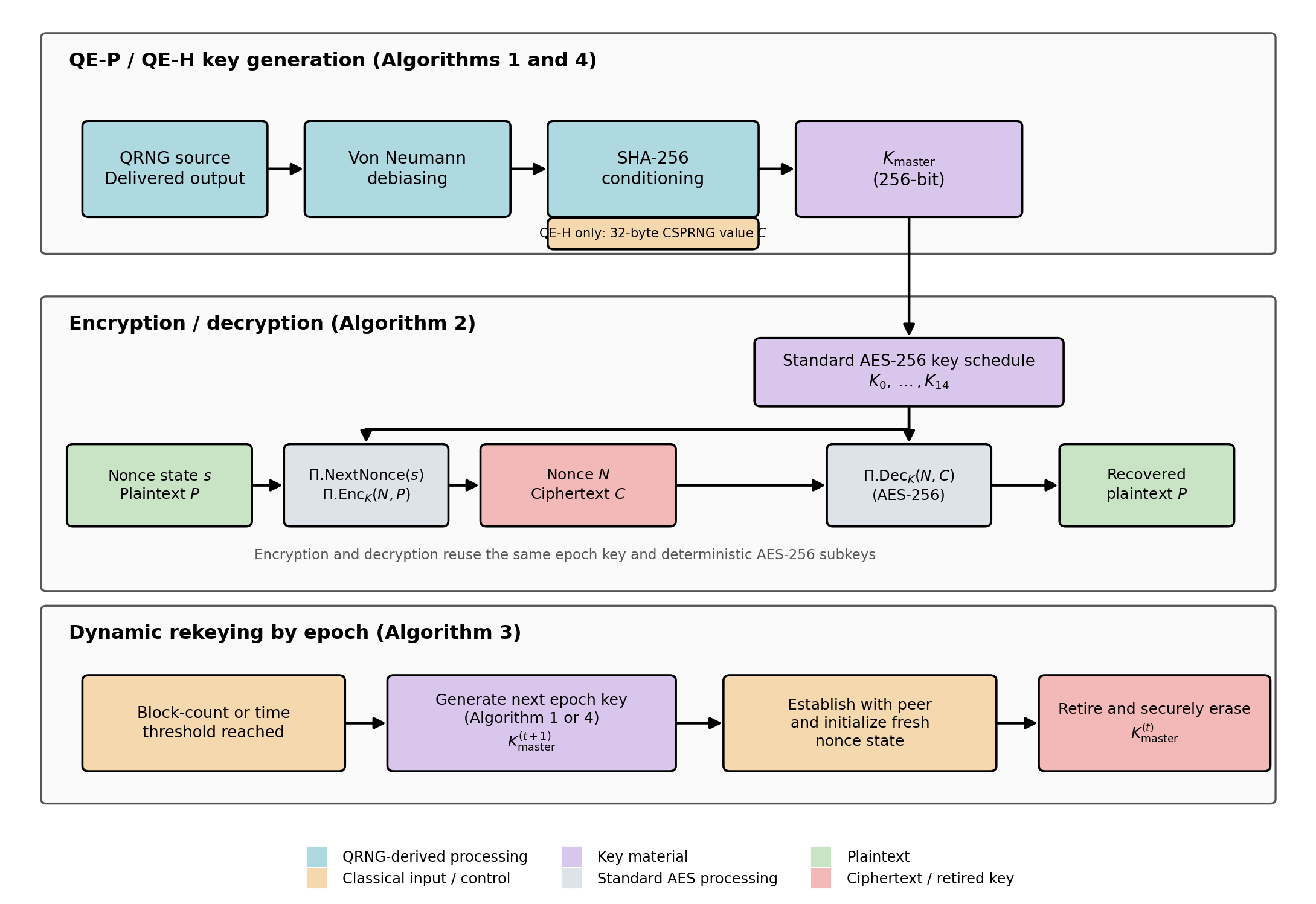}
    \caption{QE-AES system architecture. Top: key generation from delivered QRNG output (QE-H additionally mixes classical-source output). Middle: a fixed nonce-based AES-256 scheme with unique nonces under each key. Bottom: the dynamic-rekeying control loop. Secure erasure and peer establishment are deployment responsibilities.}
    \label{fig:architecture}
\end{figure}

Figure~\ref{fig:architecture} summarizes the two QE-AES key-management modes. In QE-P, delivered QRNG output is additionally debiased and conditioned into a 256-bit master key; QE-H also incorporates classical-source output. The master key drives the \emph{unmodified} AES-256 key schedule, encryption, and decryption, while a separate control loop periodically retires the current epoch key and installs a new one. Algorithms~\ref{alg:qep-keygen}--\ref{alg:qep-rekey} specify QE-P, and Algorithm~\ref{alg:qeh-keygen} specifies QE-H.

\subsection{Experimental setup}
The entropy datasets were captured from an ID Quantique Quantis USB 4M QRNG \cite{IDQuantique2025}. Performance experiments were subsequently run against the captured stream, while a separate live-device probe measured USB delivery throughput. Table~\ref{table:quantis} summarizes the device specification, and Table~\ref{table:system} identifies the capture and analysis hosts.

The Quantis USB 4M serves as the device input in our framework, with a specified delivery rate of 4 Mbit/s (with a 10\% tolerance). Unlike ID Quantique's PCIe form factor, the USB product line exposes only output processed by an internal deterministic random bit generator (DRBG) over the host interface; a raw, unconditioned detector-level tap is unavailable by design \cite{IDQuantiqueFAQ2026}. Section~\ref{sec4} therefore labels the captured stream as vendor-post-processed and evaluates only that observable boundary. The prototype applies additional Von Neumann and SHA-256 conditioning on top of the delivered output. Host-side experiments ran on an Apple M3 Pro; Table~\ref{table:system} records the separate Apple-Silicon host used for live capture.

\begin{table}[h!]
\caption{Quantis USB 4M QRNG specifications.}
\centering
\begin{tabular}{ll}
\toprule
\textbf{Specification} & \textbf{Value} \\
\midrule
Random Bit Rate & 4 Mbit/s $\pm$ 10\% \\
Thermal Noise Contribution & $<$ 1\% \\
Storage Temperature & -25°C to +85°C \\
Dimensions & 61 mm $\times$ 31 mm $\times$ 114 mm \\
USB Specification & 2.0 \\
Requirement & PC with USB connector \\
Power & Via USB port \\
\bottomrule
\end{tabular}
\label{table:quantis}
\end{table}

\begin{table}[h!]
\caption{Host environments used for capture and subsequent analysis.}
\centering
\begin{tabular}{p{0.28\textwidth}p{0.38\textwidth}p{0.22\textwidth}}
\toprule
\textbf{Role} & \textbf{Processor} & \textbf{Operating system} \\
\midrule
Analysis and benchmarks & Apple M3 Pro, 12 cores, 36 GB RAM & macOS 15.2 \\
Live-device capture & Apple Silicon (model not recorded) & macOS 26.3 \\
\bottomrule
\end{tabular}

\label{table:system}
\end{table}

\subsection{Pure quantum randomness architecture}

In the quantum-source mode (QE-P), the AES master key is derived from output delivered by the QRNG. Standard AES-256 then deterministically derives every round subkey from that master key; no independently sampled per-round value is used. The Quantis USB interface supplies a vendor-post-processed byte stream whose MSB-first bit expansion is

\begin{equation}
    \{b_i\}_{i=1}^{N}, \qquad b_i\in\{0,1\},
\end{equation}

whose physical origin and internal conditioning are vendor claims; this study evaluates only the delivered bytes.

Although the delivered stream is already conditioned inside the device, the prototype applies an additional pairwise Von Neumann debiasing stage before SHA-256 conditioning. This extra debiasing is not required for an ideal DRBG output and reduces usable throughput; on the 1\,GB capture it retained 25.01\% of the input bytes. We retain it to evaluate the originally proposed pipeline rather than as a recommended treatment of every vendor-conditioned source. Conceptually, if
\begin{equation}
(b_{2j}, b_{2j+1}) = (0,1)
\end{equation}
output ‘0’; if
\begin{equation}
(b_{2j}, b_{2j+1}) = (1,0)
\end{equation}
output ‘1’; otherwise discard. 

Let the result of this bias-correction stage be \(R\).

From \(R\), a 256-bit AES master key is produced using SHA-256 as a vetted conditioning function under NIST SP 800-90B:
\begin{equation}
K_{\text{master}} = \text{Hash}(R) \in \{0,1\}^{256}.
\end{equation}
This step condenses the input into a fixed-length key. The normative prototype behavior is to hash the full accumulated debiased buffer from the final read, rather than truncate it to exactly 256 bits; QE-H likewise hashes the full accumulated \(Q'\) followed by \(C\). It does not by itself prove exact uniformity. The assessed output entropy depends on the input min-entropy, the conditioning construction, and its implementation assumptions \cite{nist2018sp80090b}.

In the QE-P design, the classical AES-256 key schedule is applied to \(K_{\text{master}}\) \emph{without modification}, producing the standard round subkeys
\begin{equation}
K_0, K_1, K_2, \dots, K_{14},
\end{equation}
and encryption and decryption proceed exactly as in standard AES-256. We deliberately do \emph{not} inject fresh per-round quantum bits into the subkeys at encryption time. Any such per-round value would have to be reproduced by the decrypting party: if it were drawn freshly and nondeterministically from the QRNG it could never be regenerated, so decryption would fail; and if it were stored or transmitted it would constitute additional secret state requiring its own protection. Confining quantum randomness to key (and epoch) generation keeps QE-P correct by construction and bit-for-bit compatible with any conforming AES-256 implementation, library, or hardware accelerator. Freshness is instead supplied at the granularity of the \emph{key epoch} through dynamic rekeying (described below), in which the QRNG is re-invoked to produce an entirely new master key that both parties establish over their existing key-agreement channel.

Algorithms~\ref{alg:qep-keygen}--\ref{alg:qep-rekey} give the complete specification. Key generation conditions delivered source output into a master key; encryption and decryption are the unmodified AES-256 routines keyed by \(K_{\text{master}}\); and dynamic rekeying bounds the volume of data protected under any single key.

\begin{algorithm}[!htbp]
\caption{QE-P key generation}\label{alg:qep-keygen}
\begin{algorithmic}[1]
\Require Quantum source $\mathsf{QRNG}$; key length $n=256$; prototype read size $m=1024$ bytes
\Ensure Master key $K_{\text{master}}$ and round subkeys $(K_0,\dots,K_{14})$
\State $R \gets \epsilon$
\While{$|R|_{\mathrm{bits}} < n$}
  \State $\{b_i\} \gets \mathsf{QRNG.read}(m)$ \Comment{vendor-post-processed output}
  \State $R \gets R \parallel \textsc{VonNeumann}(\{b_i\})$
\EndWhile
\State $K_{\text{master}} \gets \mathrm{SHA\text{-}256}(R)$ \Comment{vetted conditioning function}
\State $(K_0,\dots,K_{14}) \gets \textsc{AES256-KeySchedule}(K_{\text{master}})$
\State \Return $K_{\text{master}},\,(K_0,\dots,K_{14})$
\end{algorithmic}
\end{algorithm}

\begin{algorithm}[!htbp]
\caption{QE-P encryption and decryption with a nonce-based AES-256 scheme}\label{alg:qep-encdec}
\begin{algorithmic}[1]
\Require Fixed nonce-based AES-256 profile $\Pi$; mutable nonce state \(s\) for the current key
\Function{Encrypt}{$K_{\text{master}}, s, P$}
  \State $(N,s') \gets \Pi.\mathsf{NextNonce}(s)$ \Comment{\(N\) must be unique under this key}
  \State $C \gets \Pi.\mathsf{Enc}_{K_{\text{master}}}(N,P)$
  \State \Return $(s',N,C)$
\EndFunction
\Function{Decrypt}{$K_{\text{master}}, N, C$}
  \State \Return $\Pi.\mathsf{Dec}_{K_{\text{master}}}(N,C)$
\EndFunction
\end{algorithmic}
\end{algorithm}

\begin{algorithm}[!htbp]
\caption{QE-P dynamic rekeying}\label{alg:qep-rekey}
\begin{algorithmic}[1]
\Require Current key $K_{\text{master}}^{(t)}$; interval $\Delta T = \min(T_{\text{block}}, T_{\text{time}})$
\State \textbf{on} trigger (block count or elapsed time reaches $\Delta T$):
    \State \quad $\big(K_{\text{master}}^{(t+1)}, \cdot\big) \gets$ \Call{QE-P.KeyGen}{} \Comment{new conditioned key}
\State \quad establish $K_{\text{master}}^{(t+1)}$ with peer over existing key-agreement channel
\State \quad initialize nonce state for epoch $t+1$ and prevent reuse under that key
\State \quad $\textsc{SecureErase}(K_{\text{master}}^{(t)})$
\end{algorithmic}
\end{algorithm}

The selected profile for \(\Pi\) must fix the AES mode, nonce width and encoding, the deterministic \(\mathsf{NextNonce}\) transition, ciphertext and tag format, and nonce-state synchronization. All communicating peers must use the same profile. Algorithm~\ref{alg:qep-rekey} initializes a fresh nonce state whenever it installs a new epoch key. Section~\ref{sec4} instantiates AES-256-CTR for bulk measurements and AES-256-GCM for the TLS-like workload; the prototype does not define a general wire protocol.

The security-relevant objective of QE-P is precise and modest: request separately provisioned epoch keys from an entropy path distinct from the host CSPRNG. Independence remains a source-model assumption. Our measurements characterize the \emph{delivered, vendor-post-processed stream}; because the Quantis USB exposes neither detector-level samples nor its internal conditioning state, they are not a certification of the underlying physical noise source. SHA-256 is a vetted SP 800-90B conditioning function, but the statistical tests reported here do not prove that its output is perfectly uniform. We stress that QE-P does not enlarge the AES-256 key space or raise the \(\Theta(2^{128})\) generic Grover cost. The Appendix states confidentiality as a function of an explicit mode-specific key-distribution defect rather than asserting that the experiments establish a negligible defect.

Because QE-P changes only key provenance and not the algorithm, rekeying limits the data and traces associated with one epoch key. Under the assumptions that epoch keys are independently generated and retired keys are securely erased, compromise of one epoch key does not directly reveal another. This is key separation, not a general claim of protocol-level forward or backward secrecy.

\subsection{Hybrid quantum-classical architecture}

In the hybrid quantum--classical mode (QE-H), AES key material is derived from delivered QRNG output \(Q'\) and an approved classical-source value \(C\). The prototype concatenates these inputs and hashes them:
\begin{equation}
K_{\text{master}} = \mathrm{SHA\text{-}256}(Q' \parallel C).
\end{equation}
We do not prove that this construction is a robust combiner or that it remains secure when one input is adversarial or predictable. QE-H is an exploratory specification and was not separately evaluated in the experiments. As in QE-P, the classical AES-256 key schedule, round function, ciphertext format, and decryption procedure remain unchanged.

\begin{algorithm}[H]
\caption{QE-H hybrid key generation}\label{alg:qeh-keygen}
\begin{algorithmic}[1]
\Require Quantum source $\mathsf{QRNG}$; approved classical source $\mathsf{CSPRNG}$; $n=256$; $m=1024$ bytes
\Ensure Master key $K_{\text{master}}$
\State $Q' \gets \epsilon$
\While{$|Q'|_{\mathrm{bits}} < n$}
  \State $Q \gets \mathsf{QRNG.read}(m)$ \Comment{vendor-post-processed output}
  \State $Q' \gets Q' \parallel \textsc{VonNeumann}(Q)$
\EndWhile
\State $C \gets \mathsf{CSPRNG.read}(32)$
\State $K_{\text{master}} \gets \mathrm{SHA\text{-}256}(Q' \parallel C)$ \Comment{hash full accumulated \(Q'\)}
\State \Return $K_{\text{master}}$
\end{algorithmic}
\end{algorithm}
QE-H uses Algorithm~\ref{alg:qep-encdec} unchanged and uses the rekeying procedure in Algorithm~\ref{alg:qep-rekey} with its key-generation call replaced by Algorithm~\ref{alg:qeh-keygen}.
\subsection{Entropy monitoring and key lifecycle}

Production deployments should continuously monitor observable source outputs and trigger corrective actions if anomalies are detected. The prototype evaluates these diagnostics offline and does not implement this continuous control loop.

A monitoring subsystem can collect periodic batches and apply frequency, correlation, and run-length diagnostics to detect selected departures from the expected source behavior.

Diagnostic bounds should be defined against the source model and applicable NIST SP 800-series guidance. If a diagnostic deviates from its accepted range, the implementation must stop key generation or switch to an approved fallback, record the event, and notify operators. Such health tests detect selected failure modes; they do not establish unpredictability when the source model itself is invalid.

A robust key lifecycle must protect key material at every stage. As specified in Algorithm~\ref{alg:qep-keygen}, \(K_{\text{master}}\) is the SHA-256-conditioned result of \(R\), or in QE-H of the concatenated inputs. This produces a 256-bit value but does not by itself prove a uniform distribution. We omit auxiliary contextual data so that the implementation matches the stated prototype.

Once AES expands \(K_{\text{master}}\) into round subkeys via the unmodified key schedule, encryption proceeds using the unmodified AES-256 round function. Rather than altering round subkeys with per-round values, the design requests a separately provisioned master key for each epoch. Independence between epoch keys is a source-model assumption, not a conclusion of the statistical tests.

To reduce exposure after compromise or leakage, the system periodically requests a new key. This may be triggered after a block count or time interval. Provided the provisioning path generates independent keys, knowledge of one epoch key does not directly determine another.

When a key is retired, the implementation invokes the platform's supported zeroization mechanism and removes references to the key. Whether physical erasure is achieved depends on the language runtime, operating system, storage medium, and cryptographic-module boundary and was not measured here.

\subsection{Deployment interfaces}

Organizations can integrate the proposed quantum-enabled AES approach with existing hardware security modules (HSMs), key management services (KMS), and cryptographic libraries such as OpenSSL through their supported entropy and key-import interfaces. HSMs are widely used to store and manage cryptographic keys within secure, tamper-resistant hardware. In a typical setup, the QRNG delivers its available output into a conditioning process, which is then passed to the HSM through an authenticated interface. The HSM may use these values to supplement an approved entropy path or to import externally generated symmetric keys, subject to the module's security policy and validation boundary.

A similar pattern applies to software-based KMS solutions, where a trusted service receives QRNG output, performs required conditioning and health checks, and exposes keys through an authenticated API. Cryptographic-library integration must use supported provider or entropy interfaces rather than assume an interchangeable callback on every version. These paths preserve the AES ciphertext format, but FIPS 140 compliance is not automatic: the complete entropy path, conditioning implementation, cryptographic module, and operational controls require validation under the applicable program.

The reference implementation includes entropy-source abstractions for the OS CSPRNG, an AES-CTR DRBG baseline, captured Quantis output, and live-device access through the vendor CLI. Entropy analyses use captured device output. The throughput, key-generation, rekeying, and TLS-like experiments use a prerecorded capture and therefore measure host-side conditioning rather than USB wait time; a separate live probe measures the device-delivery limit. The implementation and commands are available from the corresponding author upon reasonable request.

\section{Experimental results}\label{sec4}

This section reports statistical diagnostics, delivered-stream min-entropy estimates, synthetic failure tests, host-side performance measurements, a live-device throughput probe, and a simulated side-channel experiment. Section~\ref{sec5} separately discusses the security interpretation and limitations of these results.

\subsection{Delivered-stream entropy diagnostics (SP 800-90B)}

We apply SP 800-90B Most-Common-Value (MCV) estimates to the 1\,GB Quantis capture and to the same data after the prototype's Von Neumann stage \cite{nist2018sp80090b}. These stream diagnostics do not evaluate the subsequent SHA-256 key-extraction step or establish the distribution of generated keys. Table~\ref{table:entropy90b} also reports prototype Repetition Count Test (RCT) and Adaptive Proportion Test (APT) event counts. A nonzero count means that a diagnostic window crossed its configured threshold; it is not by itself a failed SP 800-90B validation. These diagnostics are evaluated offline over the delivered stream and synthetic faults and do not validate the inaccessible detector-level noise source. The near-maximal estimates for the captured stream reflect the vendor's internal DRBG post-processing. As documented by the manufacturer \cite{IDQuantiqueFAQ2026}, the Quantis USB exposes no raw detector output.

% Auto-generated by scripts/entropy_validation.py -- do not edit by hand.
\begin{table}[h!]
\caption{Delivered-stream entropy estimates and offline RCT/APT threshold events before and after Von Neumann debiasing. The captured input is vendor-post-processed USB output; these results do not certify the detector-level source or evaluate SHA-256 key outputs.}
\centering
\begin{tabular}{lcc}
\toprule
Metric & As captured & After Von Neumann \\
\midrule
Shannon entropy (bits/byte) & 7.999999 & 7.999999 \\
MCV min-entropy (bits/bit) & 1.0000 & 0.9999 \\
MCV min-entropy (bits/byte) & 7.9935 & 7.9910 \\
RCT threshold events & 54 & 17 \\
APT threshold events & 2 & 0 \\
\bottomrule
\end{tabular}
\label{table:entropy90b}
\end{table}

We additionally ran the official SP 800-90B \texttt{ea\_non\_iid} tool on a 100\,MB slice of the delivered stream. It returned \(H_{\text{original}}=7.4649\) bits/byte, more conservative than the native MCV estimate. We also ran \texttt{ea\_restart} on a \(1000\times1000\) matrix produced by restarting the host CLI process; the tool accepted the matrix and returned \(\min(H_r,H_c,H_I)=7.3538\) bits/byte. Because the device was not power-cycled and samples were not collected directly from its noise source, this is an exploratory tool result rather than a conforming restart validation or certified entropy floor.

Table~\ref{table:failuremodes} reports eight software-injected faults plus one healthy control, evaluated with fixed RCT/APT cutoffs. These diagnostics detect the injected distributional faults but not a predictable, statistically flat linear-congruential substitute.

% Auto-generated by scripts/failure_mode_eval.py -- do not edit by hand.
\begin{table}[h!]
\caption{Response of prototype RCT/APT diagnostics to software-injected failure modes using fixed baseline cutoffs. ``Event'' denotes at least one threshold crossing; these are not physical environmental trials.}
\centering
\small
\begin{tabular}{lcccc}
\toprule
Injected mode & RCT & APT & Detected & Expected \\
\midrule
Healthy baseline & No event & No event & no & no \\
Constant zero & Event & Event & yes & yes \\
Constant one & Event & Event & yes & yes \\
60\% one-bit bias & No event & Event & yes & yes \\
75\% one-bit bias & Event & Event & yes & yes \\
90\% one-bit bias & Event & Event & yes & yes \\
2\% intermittent stuck & Event & Event & yes & yes \\
Periodic pattern & No event & Event & yes & yes \\
Predictable flat LCG & No event & No event & no & no \\
\bottomrule
\end{tabular}
\label{table:failuremodes}
\end{table}

Together, the results show that health diagnostics can expose selected failures but cannot replace source assurance. Physical temperature and supply-voltage trials were not performed.

\clearpage
\subsection{Performance evaluation}

Because QE-AES uses the unmodified AES-256 core, its additional runtime cost lies in entropy delivery, conditioning, and rekeying. Tables~\ref{table:throughput}--\ref{table:tls} report host-side experiments against a prerecorded Quantis capture, the operating-system CSPRNG, and an AES-CTR DRBG. For each workload, the scheme \(\Pi\) is fixed: bulk confidentiality benchmarks use AES-256-CTR, and the TLS-like record benchmark uses AES-256-GCM; nonce/counter state is not reused under an epoch key. The experiments isolate conditioning and cipher costs but exclude live USB waiting time. A separate live-device probe supplies the deployment throughput bound.

\begin{itemize}
    \item \textbf{Bulk file / cloud-object-storage encryption.} A single 100\,MB object (representative of a file- or blob-storage upload) encrypted under dynamic rekeying at intervals $\Delta T \in \{1,10,100\}$\,MB, as reported in Table~\ref{table:rekeying}.
    \item \textbf{TLS-like session encryption.} A 50\,MB session transmitted as $16$\,KB AES-256-GCM records (the maximum TLS record size) with a capture-backed Quantis-source rekey every $1000$ records---analogous to a TLS~1.3 \texttt{KeyUpdate}, but on our epoch schedule rather than triggered by the peer. Because per-record latency, not aggregate throughput, is the operationally relevant metric for this workload, Table~\ref{table:tls} reports latency percentiles alongside throughput, comparing the capture-backed Quantis pipeline with direct 256-bit OS CSPRNG rekeying.
\end{itemize}

% Auto-generated by scripts/benchmark_qeaes.py -- do not edit by hand.
\begin{table}[h!]
\caption{AES-256-CTR encryption throughput by payload size (QE-AES uses the identical cipher core; the difference is in key provisioning, quantified separately).}
\centering
\begin{tabular}{lr}
\toprule
Payload & Throughput (MB/s) \\
\midrule
1 KB & 149.3 \\
1 MB & 216.5 \\
10 MB & 194.9 \\
100 MB & 193.7 \\
1 GB & 195.2 \\
\bottomrule
\end{tabular}
\label{table:throughput}
\end{table}

% Auto-generated by scripts/benchmark_qeaes.py -- do not edit by hand.
\begin{table}[h!]
\caption{Capture-backed host-side conditioning latency and CPU cost for a 256-bit key, compared with direct 256-bit OS CSPRNG and AES-CTR DRBG output. The capture row excludes live USB delivery. CPU (ms) is process CPU time per key (user+sys).}
\centering
\begin{tabular}{lrrrr}
\toprule
Source & Mean (ms) & Median (ms) & Keys/s & CPU (ms) \\
\midrule
OS CSPRNG & 0.001 & 0.001 & 795949 & 0.0014 \\
AES-CTR DRBG & 0.002 & 0.001 & 483780 & 0.0021 \\
QRNG capture & 0.028 & 0.024 & 36324 & 0.0269 \\
\bottomrule
\end{tabular}
\label{table:keygen}
\end{table}

% Auto-generated by scripts/benchmark_qeaes.py -- do not edit by hand.
\begin{table}[h!]
\caption{Capture-backed AES-256-CTR rekeying performance from prerecorded Quantis output compared with single-key throughput; live USB delivery is excluded.}
\centering
\begin{tabular}{rrrr}
\toprule
Rekey interval $\Delta T$ & Rekeys & Throughput (MB/s) & Overhead vs.\ single key \\
\midrule
1 MB & 100 & 190.0 & 1.9\% \\
10 MB & 10 & 192.9 & 0.4\% \\
100 MB & 1 & 196.6 & $-$1.5\% \\
\bottomrule
\end{tabular}
\label{table:rekeying}
\end{table}

% Auto-generated by scripts/benchmark_qeaes.py -- do not edit by hand.
\begin{table}[h!]
\caption{AES-256-GCM TLS-like microbenchmark (16384-byte records; rekey every 1000 records; 3 rekeys). The capture-backed Quantis arm uses the prototype conditioning pipeline; the OS baseline uses direct 256-bit CSPRNG output. Live USB delivery is excluded.}
\centering
\small
\begin{tabular}{lrrrrr}
\toprule
Rekey source & Throughput (MB/s) & Records/s & p50 (ms) & p95 (ms) & p99 (ms) \\
\midrule
QRNG capture & 84.7 & 5421 & 0.1536 & 0.2672 & 0.3820 \\
OS CSPRNG & 77.2 & 4943 & 0.1547 & 0.2902 & 0.6042 \\
\bottomrule
\end{tabular}
\label{table:tls}
\end{table}

The capture-backed results show that AES-256 throughput is independent of key provenance and that host-side conditioning costs about \(0.028\) ms per key. Peak resident memory for the full harness was 73.4\,MB and was dominated by payload buffering. These figures must not be interpreted as live QRNG provisioning latency. The separate live probe delivered 3.74\,Mbit/s; with the prototype's 1024-byte initial read per key (and additional reads only if debiasing yields fewer than 256 bits), this corresponds to at most approximately 479 keys/s, or at least 2.1\,ms of entropy-delivery time per key. The small differences in Table~\ref{table:rekeying}, including the \(-1.5\%\) value at the 100\,MB interval, reflect run-to-run benchmark variation rather than a rekeying speedup and describe capture-backed execution only. The TLS-like CSPRNG arm uses direct 256-bit OS output rather than the Quantis conditioning path. At aggressive intervals such as 1\,MB, live USB delivery can be a material overhead and requires an asynchronous entropy pool or a less frequent policy. The TLS-like run contains only three rekeys and does not establish tail latency for a live-device rekey event.

\subsection{Side-channel analysis: trace limiting}\label{sec:sidechannel}

To make the trace-limiting argument concrete---and to answer explicitly whether QE-AES behaves as a masking technique, a rekeying strategy, or merely an entropy improvement (it is a rekeying strategy)---we simulate a correlation power analysis (CPA) attack under the textbook Hamming-weight leakage model. For a target key byte, we model the leakage of the first-round S-box output as $L = \mathrm{HW}(\mathrm{SBox}(p \oplus k)) + \mathcal{N}(0,\sigma^2)$ and recover the key by correlating hypothesized leakage with observed leakage over many traces. The security-relevant quantity is the \emph{traces-to-disclosure} (MTD): the number of measurements, all collected under the \emph{same} key, required for recovery. Table~\ref{table:sidechannel} reports the MTD across signal-to-noise ratios, and Figure~\ref{fig:sidechannel} shows the key-recovery success rate as a function of the number of traces available per key.

A static-key baseline permits modeled traces to accumulate under one key. MTD is expressed in comparable leakage traces per key, whereas the performance experiments specify rekeying in bytes or records. Translating between them requires a deployment-specific mapping from protected operations to obtainable traces; this study does not claim that any benchmark interval satisfies the simulated MTD threshold. The experiment covers one AES key byte under synthetic Hamming-weight leakage and Gaussian noise, not physical timing, power, electromagnetic, cache, profiling, or multi-epoch attacks.

% Auto-generated by scripts/sidechannel_sim.py -- do not edit by hand.
\begin{table}[h!]
\caption{Simulated CPA disclosure threshold (MTD) for one AES key byte under a Hamming-weight leakage model. MTD is a modeled trace count, not a byte-, block-, or record-based rekeying interval. This is a simulation, not measured leakage.}
\centering
\begin{tabular}{rr}
\toprule
SNR & MTD (modeled traces/key) \\
\midrule
2.000 & 20 \\
0.500 & 80 \\
0.222 & 160 \\
0.125 & 320 \\
0.080 & 320 \\
\bottomrule
\end{tabular}
\label{table:sidechannel}
\end{table}

\IfFileExists{img/sidechannel_mtd.png}{%
  \begin{figure}[h!]
  \centering
  \includegraphics[width=0.7\textwidth]{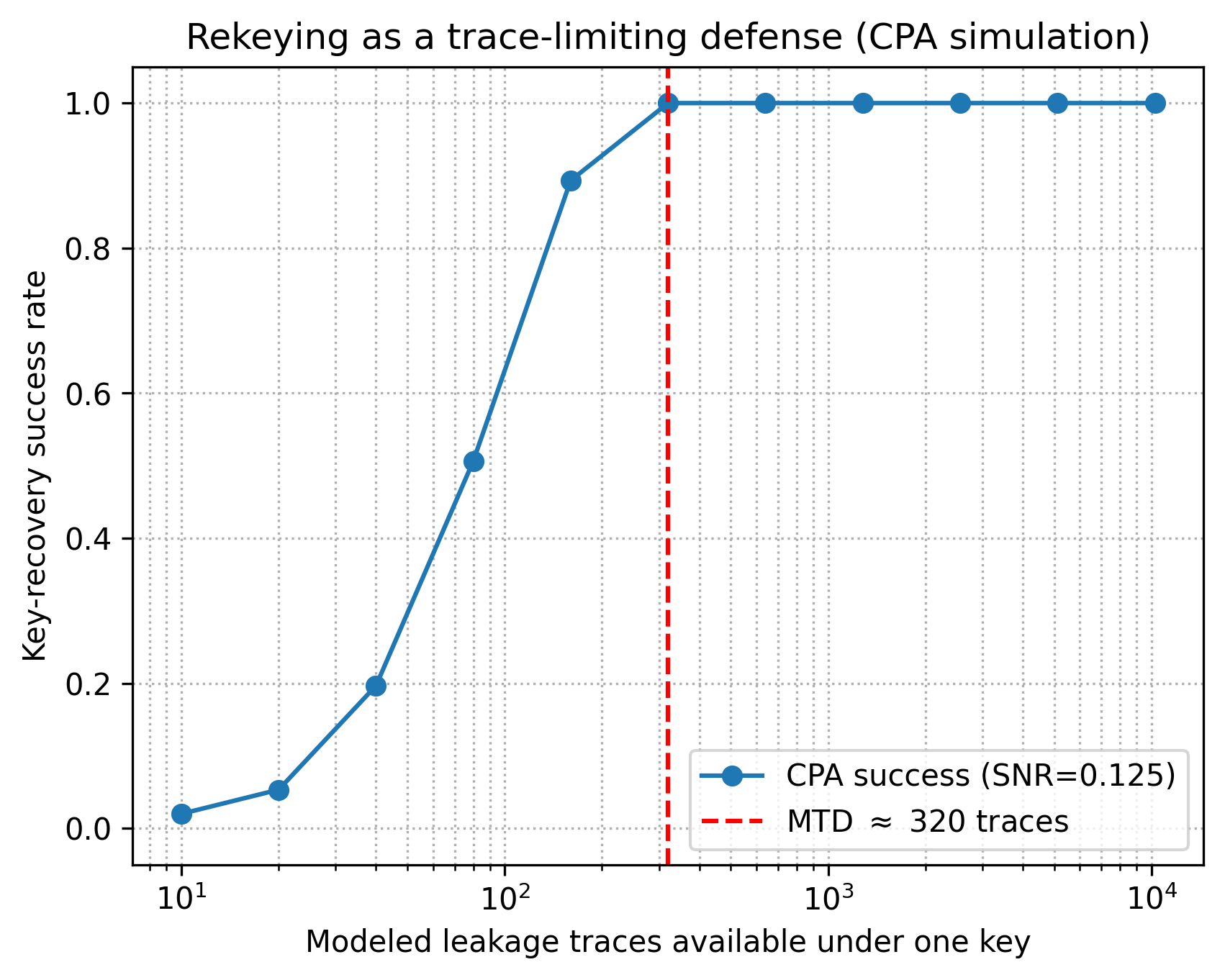}
  \caption{Simulated CPA key-recovery success rate at SNR \(=0.125\) versus modeled leakage traces available under one key, illustrating the trace budget that a rekeying policy would need to limit. The dashed line marks the model's traces-to-disclosure (MTD), not a byte-, block-, or record-based rekeying interval. The curve is independent of whether the epoch key is provisioned by a QRNG or a classical source.}
  \label{fig:sidechannel}
  \end{figure}%
}{}

\clearpage
\section{Discussion and limitations}\label{sec5}

\subsection{Security interpretation and deployment scope}
For a 256-bit key, generic Grover search has complexity on the order of \(2^{128}\), and no method of generating the key raises that asymptotic bound. A separately engineered entropy path can reduce dependence on a host CSPRNG, but this study does not prove that the resulting key distribution is perfectly uniform or cryptographically superior to a correctly implemented CSPRNG.

Rekeying limits the amount of ciphertext and the number of traces associated with one epoch key. This benefit is not unique to quantum-derived keys: classical independent rekeying provides the same lifetime bound. QRNG input changes source diversity, while the policy supplies trace limiting. Neither mechanism substitutes for authenticated key establishment, nonce management, masking, or other implementation countermeasures.

Table~\ref{table:threats} separates properties supported by the design from properties that remain assumptions or deployment responsibilities. The statistical diagnostics found no systematic anomaly in the tested delivered streams; they do not certify the physical source or prove key uniformity.

\begin{table}[h!]
\caption{Scope of QE-AES claims and remaining deployment requirements.}
\centering
\small
\begin{tabular}{l l}
\toprule
Threat vector               & Supported statement                                \\
\midrule
Grover's algorithm          & Generic AES-256 bound unchanged (\(\approx2^{128}\)) \\
Key compromise or HNDL      & Rekeying limits data protected by one epoch key \\
Side-channel accumulation   & Trace count is policy-dependent; no masking provided \\
Host CSPRNG compromise      & Separate device input path; host isolation not evaluated \\
\bottomrule
\end{tabular}
\label{table:threats}
\end{table}

From a deployment perspective, potential applications include file encryption, cloud-object storage, and session-key management where an independent entropy source is operationally justified. Adoption requires an authenticated key-establishment channel, atomic epoch transitions, nonce separation across keys, secure erasure, entropy buffering, failure handling, and validation within the HSM or KMS security boundary. These controls were not implemented or benchmarked in this study. PQC-based key establishment and separately provisioned AES epoch keys address distinct lifecycle stages: PQC protects establishment of shared secrets, whereas QE-AES concerns endpoint key generation and rotation. Quantum key distribution is another possible establishment mechanism under its own assumptions; it is not a property of AES-256 or QE-AES.

\subsection{Limitations and future work}\label{sec:future}

The principal limitations are the Quantis USB's inaccessible detector-level output, the use of CLI-process rather than physical-device restarts, and the absence of temperature and supply-voltage trials. RCT/APT diagnostics were evaluated offline rather than as a continuously deployed streaming monitor. The performance tables use prerecorded data; only the 3.74\,Mbit/s probe measures live USB delivery. The TLS-like workload performs three capture-backed rekeys and is not a complete TLS implementation, and the bulk-rekeying sweep has no matched CSPRNG-rekeying arm. The CPA experiment is simulated and covers one byte under one leakage model. HSM/KMS integration, QE-H combiner security and performance, and physical secure-erasure behavior were not evaluated.

Future work should obtain detector-level samples from suitable hardware, perform conforming physical restart and environmental tests, deploy streaming health tests, and benchmark live asynchronous entropy pools under realistic rekey policies with matched QRNG and CSPRNG rekeying arms. Physical power and electromagnetic measurements should compare standard AES-256 and QE-AES directly. A protocol implementation should specify epoch identifiers, key confirmation, rollback behavior, and nonce management.

\section{Conclusion}\label{sec6}

QE-AES is an engineering wrapper around unmodified AES-256, not a new cipher or an increase in Grover resistance. The study provides decryptable QE-P and QE-H key-generation specifications, measurements of a commercial QRNG's delivered stream, capture-backed performance benchmarks, a live-device throughput bound, and a scoped trace-limiting simulation. The results support source diversity and operational rekeying as deployment objectives, while also showing that USB entropy delivery can dominate aggressive rekeying. Statistical tests do not certify the inaccessible physical noise source, and rekeying does not provide masking or protocol security by itself. The framework is therefore best understood as a measured key-management design whose security depends on independently validated source and conditioning assumptions, authenticated key establishment, and conventional implementation controls.

\section*{Statements and declarations}
\textbf{Funding.} No external funding was received for this study.

\textbf{Competing interests.} The authors are employees of PricewaterhouseCoopers. The authors declare no other financial or non-financial interests directly related to this work.

\textbf{Author contributions.} A.N.M., A.K.B., and J.J.V. jointly contributed to study conception and design, material preparation, data collection, software development, analysis, and manuscript revision. All authors read and approved the final manuscript.

\textbf{Data and code availability.} The data, source code, generated results, provenance records, and reproduction instructions supporting this study are available from the corresponding author upon reasonable request. The package supplied on request provides large captures as reconstructable chunks with manifests and SHA-256 hashes for integrity verification.

\bibliography{sn-bibliography}% common bib file
\newpage
\begin{appendices}
\setcounter{table}{0}
\renewcommand{\thetable}{A\arabic{table}}
\setcounter{equation}{0}
\renewcommand{\theequation}{A\arabic{equation}}

\section*{Security analysis}\label{app:proofs}
\addcontentsline{toc}{section}{Security analysis}
This appendix makes precise what QE-AES does and does not provide. We treat QE-AES as a \emph{key-management wrapper} around AES-256 rather than as a new keyed primitive: the round function, mode of operation, and ciphertext format are those of a fixed standard AES-256 scheme, and only the provenance of the key is changed. Accordingly, we do not claim a new pseudorandom permutation or hardness assumption. For either fixed provisioning mode \(M\), we show that the confidentiality advantage differs from the uniform-key scheme by at most the mode-specific key-distribution defect \(\Delta_K^{M}\); this work does not establish that the defect is negligible.

\subsection*{Scheme definition}

We define $\mathsf{QE\text{-}AES} = (\mathsf{KeyGen}, \mathsf{Enc}, \mathsf{Dec})$ for a fixed nonce-based AES-256 encryption scheme $\Pi$:
\begin{itemize}
  \item $\mathsf{KeyGen}$: obtain delivered source output, apply the specified conditioning pipeline, and produce \(K\in\{0,1\}^{256}\) using SHA-256 (Algorithm~\ref{alg:qep-keygen}). In QE-H, the conditioned quantum input is combined with an approved classical source as in Algorithm~\ref{alg:qeh-keygen}.
  \item $\mathsf{Enc}_K = \Pi.\mathsf{Enc}_K$ and $\mathsf{Dec}_K = \Pi.\mathsf{Dec}_K$: the unmodified AES-256 mode.
\end{itemize}
Thus QE-AES and $\Pi$ differ only in how $K$ is sampled.

\subsection*{Key-distribution model}

Fix a provisioning mode \(M\in\{\mathrm{QE\text{-}P},\mathrm{QE\text{-}H}\}\). Let \(D_K^{M}\) be the distribution of keys output by that complete provisioning path and let \(U_{256}\) be uniform on \(\{0,1\}^{256}\). For distributions \(P,Q\) on a finite set, define statistical distance as
\(\mathrm{SD}(P,Q)=\tfrac12\sum_x|P(x)-Q(x)|=\max_S|P(S)-Q(S)|\).
The mode-specific key-distribution defect is
\begin{equation}
  \Delta_K^{M}=\mathrm{SD}(D_K^{M},U_{256}).
\end{equation}
This definition makes no assumption that either mode's defect is negligible. In particular, the QE-H construction has its own distribution and is not assigned the QE-P defect. SHA-256 is a vetted conditioning function under SP 800-90B, but the empirical estimators in Section~\ref{sec4} do not calculate statistical distance and do not justify a leftover-hash-lemma claim for this unseeded construction. Establishing a concrete bound on \(\Delta_K^{M}\) requires a validated model of the complete entropy source and conditioning boundary for mode \(M\), which the Quantis USB interface does not expose. The theorem below therefore states exactly how any externally established mode-specific defect enters confidentiality.

\subsection*{Security model and assumptions}

We consider a probabilistic adversary $A$ that plays the nonce-respecting IND-CPA game with access to a left-or-right encryption oracle and additionally possesses a quantum computer (and may run Grover's algorithm offline). Nonces are required not to repeat under a key. We restrict $A$ to \emph{classical} oracle queries; this is the post-quantum (``Q1'') setting. We make no integrity or nonce-misuse-resistance claim and do \emph{not} claim security against superposition (``Q2''/QS2) queries, as our construction does not modify AES.

Under Grover's algorithm, generic search over a uniform \(n\)-bit key space requires \(O(2^{n/2})\) oracle evaluations; for \(n=256\) this is \(O(2^{128})\). Key generation cannot raise this exponent. A nonuniform key distribution may reduce search cost, but this study does not derive that cost from its statistical tests. We therefore make no ``\(2^{128+\delta}\)'' or guaranteed-full-entropy claim.

\subsection*{Confidentiality theorem}

For either scheme, define IND-CPA advantage as
\(\mathrm{Adv}^{\mathrm{IND\text{-}CPA}}(A)
=\left|\Pr[A\text{ wins}]-\tfrac{1}{2}\right|\).

\begin{theorem}\label{thm:qeaes-indcpa}
Let \(\Pi\) be a nonce-respecting IND-CPA-secure encryption scheme based on AES-256 and keyed by a uniform key, and fix QE-AES mode \(M\) with key distribution \(D_K^{M}\). Then for every classical-query, nonce-respecting IND-CPA adversary \(A\) against mode \(M\) there is an adversary \(B\) against \(\Pi\), with essentially the same running time and query count, such that
\begin{equation}
  \mathrm{Adv}_{\mathsf{QE\text{-}AES}^{M}}^{\mathrm{IND\text{-}CPA}}(A)
  \;\le\;
  \mathrm{Adv}_{\Pi}^{\mathrm{IND\text{-}CPA}}(B)
  \;+\;
  \Delta_K^{M}.
\end{equation}
\end{theorem}

\begin{proof}
We use a two-game hybrid.

\emph{Game \(0\)} is the real IND-CPA game for mode \(M\): \(K\leftarrow D_K^{M}\), and \(A\) interacts with the left-or-right oracle instantiated with \(\Pi.\mathsf{Enc}_K\).

\emph{Game \(1\)} is identical except that \(K^{\ast}\leftarrow U_{256}\). Since only the key distribution changes, the maximal-event characterization of statistical distance gives
\begin{equation}
  \bigl|\Pr[A \text{ wins } G_0] - \Pr[A \text{ wins } G_1]\bigr| \;\le\; \Delta_K^{M}.
\end{equation}

Game $1$ is exactly the IND-CPA game for $\Pi$ under a uniform key; define $B$ to run $A$ against its own $\Pi$ oracle and output $A$'s guess. Thus \(\left|\Pr[A \text{ wins } G_1] - \tfrac{1}{2}\right| = \mathrm{Adv}_{\Pi}^{\mathrm{IND\text{-}CPA}}(B)\). The triangle inequality then gives the claim.
\end{proof}

Theorem~\ref{thm:qeaes-indcpa} isolates the effect of nonuniform key provisioning. It does not prove that either mode has negligible \(\Delta_K^{M}\), nor that QE-AES improves on a correctly implemented CSPRNG. The empirical work in Section~\ref{sec4} diagnoses delivered-stream behavior but is not a proof of \(\Delta_K^{M}\). Subject to an independently validated negligible mode-specific defect and the assumed post-quantum security of \(\Pi\) for classical queries, the wrapper inherits the corresponding IND-CPA bound.

For completeness, define the \(E\)-key game as one in which a single hidden challenge bit is shared by \(E\) nonce-respecting left-or-right oracles, each instantiated with an independently generated key from the fixed mode \(M\). Let \(\mathrm{Adv}_{\mathsf{QE\text{-}AES}^{M},E}\) and \(\mathrm{Adv}_{\Pi,E}\) denote advantage in this game with mode-\(M\) and uniform keys, respectively.
\begin{proposition}\label{prop:qeaes-multikey}
For every adversary \(A\) in the \(E\)-key game, there is an adversary \(B\) in the corresponding uniform-key game such that
\[
\mathrm{Adv}_{\mathsf{QE\text{-}AES}^{M},E}(A)
\leq \mathrm{Adv}_{\Pi,E}(B)+E\Delta_K^{M}.
\]
\end{proposition}
\begin{proof}
Define hybrids \(H_j\), \(0\leq j\leq E\), in which the first \(j\) oracle keys are uniform and the remaining keys follow \(D_K^{M}\). Conditioned on all other keys and randomness, adjacent hybrids differ only in one key draw, so their winning probabilities differ by at most \(\Delta_K^{M}\). Summing the \(E\) adjacent differences and applying the triangle inequality yields the stated bound; \(H_E\) is the uniform \(E\)-key game.
\end{proof}
This joint-game loss depends on independent epoch keys and does not follow from the reported statistical tests. For ciphertext protected under one specified epoch key, Theorem~\ref{thm:qeaes-indcpa} gives the per-key bound with a single \(\Delta_K^{M}\) term.

\subsection*{Rekeying scope and failure conditions}

Let \(\Delta T\) be the rekeying interval (either time-based or block-based):
\begin{equation}
  \Delta T \;=\; \min\bigl(T_{\text{block}}, T_{\text{time}}\bigr).
\end{equation}
Each time \(\Delta T\) is reached, the system requests a new epoch key (Algorithm~\ref{alg:qep-rekey}). Under an independent-key assumption, this practice:
\begin{itemize}
  \item Splits data streams across multiple encryption keys, bounding the ciphertext volume associated with any one key in a harvest-now-decrypt-later setting.
  \item Requires separate key recovery for separate epochs; the total cost scales with the number of targeted keys but the per-key Grover exponent is unchanged.
\end{itemize}

In practical terms, confidentiality can fail through:

\begin{itemize}
  \item a weakness or misuse of the selected AES-256 mode;
  \item prediction or bias in the complete key-provisioning path;
  \item compromise of key establishment, endpoint storage, nonces, or implementation leakage.
\end{itemize}

Rekeying limits exposure per key but does not repair any of these failures. Its effect is operational and is available with either quantum-derived or classical independent epoch keys.

\section*{List of acronyms}
\addcontentsline{toc}{section}{List of acronyms}%
\label{sec:acronyms}

\footnotesize
\noindent
\begin{minipage}[t]{0.48\textwidth}
\raggedright
\textbf{AES}\quad Advanced Encryption Standard\par
\textbf{APT}\quad Adaptive Proportion Test\par
\textbf{API}\quad Application Programming Interface\par
\textbf{CLI}\quad Command-Line Interface\par
\textbf{CPA}\quad Correlation Power Analysis\par
\textbf{CSPRNG}\quad Cryptographically Secure Pseudorandom Number Generator\par
\textbf{DRBG}\quad Deterministic Random Bit Generator\par
\textbf{FIPS}\quad Federal Information Processing Standards\par
\textbf{GCM}\quad Galois/Counter Mode\par
\textbf{HNDL}\quad Harvest-Now Decrypt-Later\par
\textbf{HSM}\quad Hardware Security Module\par
\textbf{IND-CPA}\quad Indistinguishability under Chosen Plaintext Attack\par
\textbf{KMS}\quad Key Management Service\par
\textbf{MCV}\quad Most Common Value\par
\end{minipage}\hfill
\begin{minipage}[t]{0.48\textwidth}
\raggedright
\textbf{MTD}\quad Measurements (or Traces) to Disclosure\par
\textbf{NIST}\quad National Institute of Standards and Technology\par
\textbf{PQC}\quad Post-Quantum Cryptography\par
\textbf{Q1}\quad Quantum adversary with classical oracle queries\par
\textbf{Q2/QS2}\quad Quantum adversary with superposition oracle queries\par
\textbf{QE-AES}\quad Quantum-Enabled AES\par
\textbf{QE-H}\quad QE-AES Hybrid Mode\par
\textbf{QE-P}\quad QE-AES QRNG-Source Mode\par
\textbf{QKD}\quad Quantum Key Distribution\par
\textbf{QRNG}\quad Quantum Random Number Generator\par
\textbf{RCT}\quad Repetition Count Test\par
\textbf{SHA}\quad Secure Hash Algorithm\par
\textbf{SNR}\quad Signal-to-Noise Ratio\par
\textbf{TLS}\quad Transport Layer Security\par
\textbf{USB}\quad Universal Serial Bus\par
\end{minipage}

\end{appendices}
\end{document}